\documentclass[a4paper]{jpconf}
\usepackage{graphicx}

\usepackage[percent]{overpic}
\usepackage{subfigure}

\def\bcor{b_{\rm corr}}
\def\av#1{\langle #1 \rangle}

\def\nF{n_{\rm F}}
\def\nB{n_{\rm B}}
\def\pt{p_{\rm T}}

\def\avr#1#2{\langle {#1} \rangle^{}_{#2}}

\begin{document}
\title{Forward-backward correlations with strange particles in PYTHIA }

\author{
I.G.Altsybeev$^1$, G.A.Feofilov$^1$, E. L. Gillies$^2$\\
{\it \small $^1$ Saint-Petersburg State University, RU}\\
{\it \small $^2$ University of Edinburgh, UK}
}


\ead{Igor.Altsybeev@cern.ch}

\begin{abstract}
We present studies of strange particle yields and correlations in $pp$ collisions in the PYTHIA8 
event generator by studying forward-backward correlations.  Several key processes that give
rise to these correlative effects are identified and manipulated to probe the fundamental properties
of strange particle emitting sources.  The sensitivity of strange particle production and 
correlations to PYTHIA's multiparton interaction, color reconnection,  and explicit strangeness
suppression are shown.  
\end{abstract}

\vspace{-0.5cm}
\section{Introduction}

Forward-backward (FB) correlations are a powerful tool for
studying the initial stages of $pp$ and AA collisions.
FB correlations are studied in two intervals
of pseudorapidity ($\eta$)
which are selected one in the forward 
and another in the backward 
hemispheres in the center-of-mass system.

The FB correlation strength is characterized by the correlation coefficient $\bcor$.
This value is defined via a linear regression of the average value
of a given quantity $B$ measured in the backward hemisphere ($\avr B F$) 
as a function of value of another quantity $F$ measured in the forward hemisphere.
Note that $F$ and $B$ can describe the kinematic quantity or distinct quantities.  
\begin{equation}\label{linearFB}
\avr B F = a + \bcor\cdot F \  .
\end{equation}

Taking $F$ and $B$ particle multiplicities, the relation 
(\ref{linearFB}) becomes
$\avr\nB\nF = a + \bcor\cdot\nF$,  
which was first experimentally observed in UA5 \cite{exp1} 
and discussed in \cite{theor1,theor2,theor3}.  FB correlations between multiplicities have been
recently studied in $pp$ and Au-Au collisions by  STAR \cite{FB_STAR} at RHIC, and in $pp$
collisions by ATLAS \cite{FB_ATLAS} and ALICE \cite{FB_JHEP} at LHC.

FB correlation studies are more informative when decoupled into short-range and long-range
components \cite{exp2, theor3}.
Short-range correlations (SRC) are localized over a small range of $\eta$, typically up to one unit. 
They are induced by various short-range effects like decays of clusters or resonances, jet and
mini-jet induced correlations.  Long-range correlations (LRC) extend over a wider range in $\eta$
and originate from fluctuations in the number and properties of particle emitting sources, e.g.
clusters, cut pomerons, strings, mini-jets etc.  \cite{theor3, exp2, two_stage_scenario,lrc_additional_1,lrc_additional_2}.
In ALICE paper \cite{FB_JHEP}, the ``classical" approach to the long-range correlation analysis in
two pseudorapidity intervals was expanded using additional azimuthal ($\varphi$) sectors within
these windows. 
This approach allows for a more thorough investigation of the SRC and LRC and their contributors,
which can provide stronger constrains on phenomenological string models.  
Correlations with additional azimuthal segmentation of rapidity windows were also studied in PYTHIA6
\cite{Skands_twisted_corrs}.


\section{Motivation for other variables in FB correlations}

FB multiplicity correlations in $pp$ collisions can be interpreted  using the parametric string
model \cite{VV_string_model} which implies event-by-event fluctuations in number of strings as 
{\it independent} particle emitters.
However, independent emitters can not describe other types of correlations, such as a non-zero
correlation between 
charged particle multiplicity and average transverse momentum ($\av{\pt}_{N_{\rm ch}}$  correlation)
in a {\it single} $\eta$-window This was first established at ISR energies in \cite{pT_N_ISR}.

The correlation of the mean $\pt$ of charged particles and other observables 
can be  explained via
collective effects relevant to the formation of particle emitting sources.  In $pp$ and $p\bar{p}$
collisions, these collective effects were considered to be string fusion between quark-gluon strings
\cite{EPEM, EPEM-2}. Specifically, the multi-Pomeron exchange model provided a description of
the experimentally measured growth in $\pt$ with event multiplicity over a wide energy range of
collision energies (0.3-1.8 TeV).   
It was shown \cite{P-color} that the use of color reconnection in string-based PYTHIA model can produce the positive $\pt$-multiplicity correlation seen experimentally in pp collisions.

In string-based models,
the Schwinger-like mechanism of string hadronization is used to obtain the production rate of
$q\overline{q}$ pairs with opposite transverse momenta $\pt$.  The rate is proportional to
$\exp{\Big(-{\pi \over \kappa}(m^2+\pt^2)􏰒}\Big)$, where $\kappa$ is a string tension and $m$ is a
quark mass.  This result can be used to estimate the relative production
of different flavoured quarks and the $\pt$ distribution.  Collective effects could yield
a higher effective string tension, as in the string fusion model \cite{SFM1, SFM2, SFM3_VK} and the
overlapping color ropes model in DIPSY event generator \cite{effects_of_overlapping_strings}.
%
Larger string tension implies larger strangeness and baryon fractions as expected.


PYTHIA allows for {\it multiple parton interactions} (MPI) in $pp$ events.
This can cause non-negligible phase-space overlaps between final states from different MPI systems.
The interaction between strings is implemented by color reconnection (CR), as proposed in \cite{CR_1996}. 
In PYTHIA 8 before reconnection, partons are connected in their respective MPI system.
The color flow of two such systems can be fused such that the partons of the lower-$\pt$ system are
added to the strings defined by the higher-$\pt$ system to give the smallest total string length.
This is the default method in PYTHIA8.  
In the new CR model \cite{new_CR}, junction structures are introduced in addition to the more common
string-string reconnections.  The new model has been was tuned to reproduce 
measured ratios of kaons and hyperons such as $\Lambda/K^0_{\rm S}$ ratio.  The use of junction
structures introduces a slight enhancement of the strangeness and overall baryon production in this
implementation.

\section{Forward-backward correlations of strange particles in PYTHIA8} 

\begin{figure}
\subfigure[Subfigure 1 list of figures text][]
{
\begin{overpic}[width=0.49\textwidth]{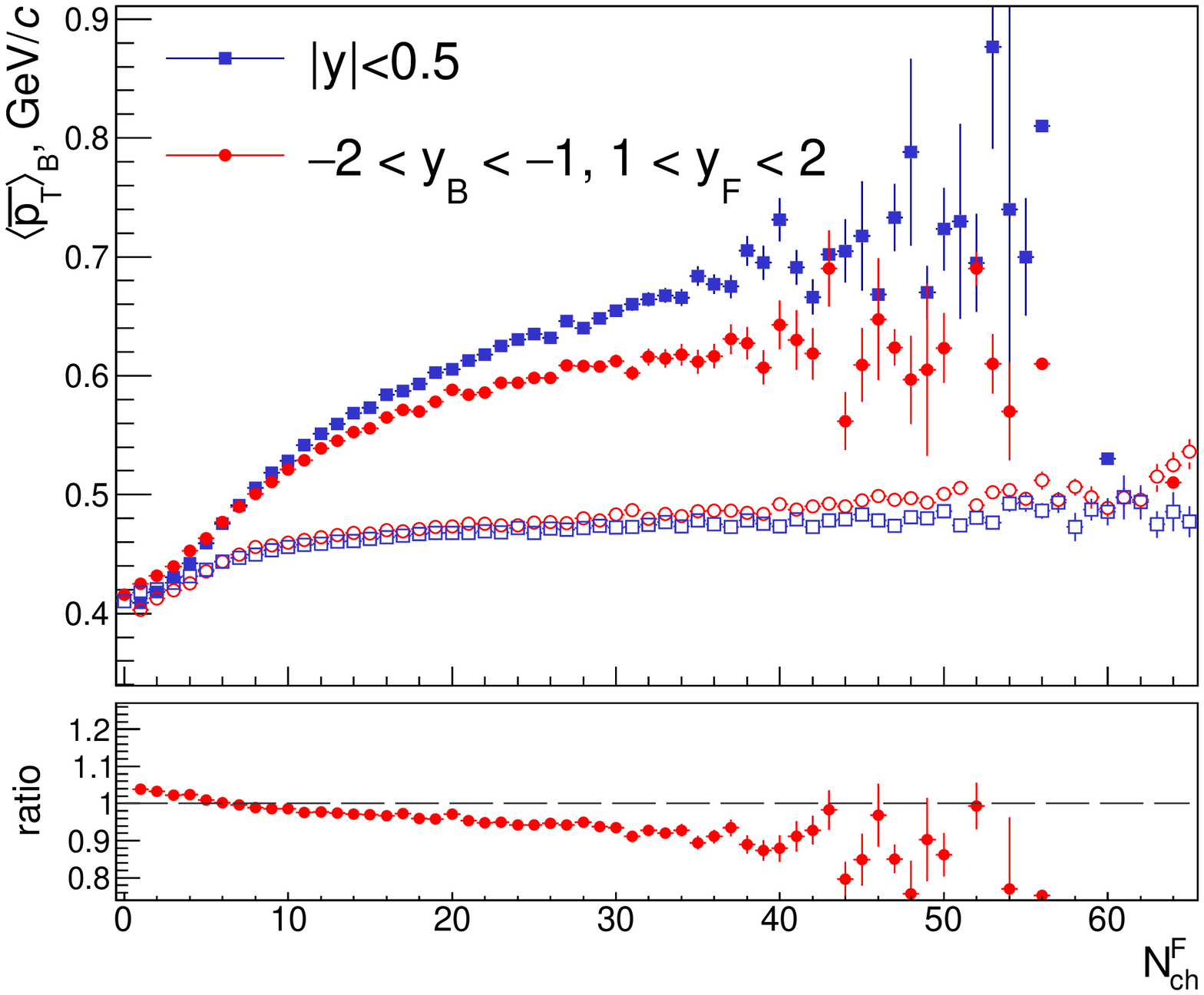} 
 \put(14,60){\small PYTHIA8 }
 \put(14,55){\small  Monash 2013}
 \put(48,47){\small CR on}
 \put(74,33){\small CR off}
 \put(66,19){\small red/blue {\tiny (CR on)} }
\end{overpic}
}
\hfill
\hspace{-0.4cm}
\subfigure[Subfigure 2 list of figures text][]
{
\begin{overpic}[width=0.49\linewidth]{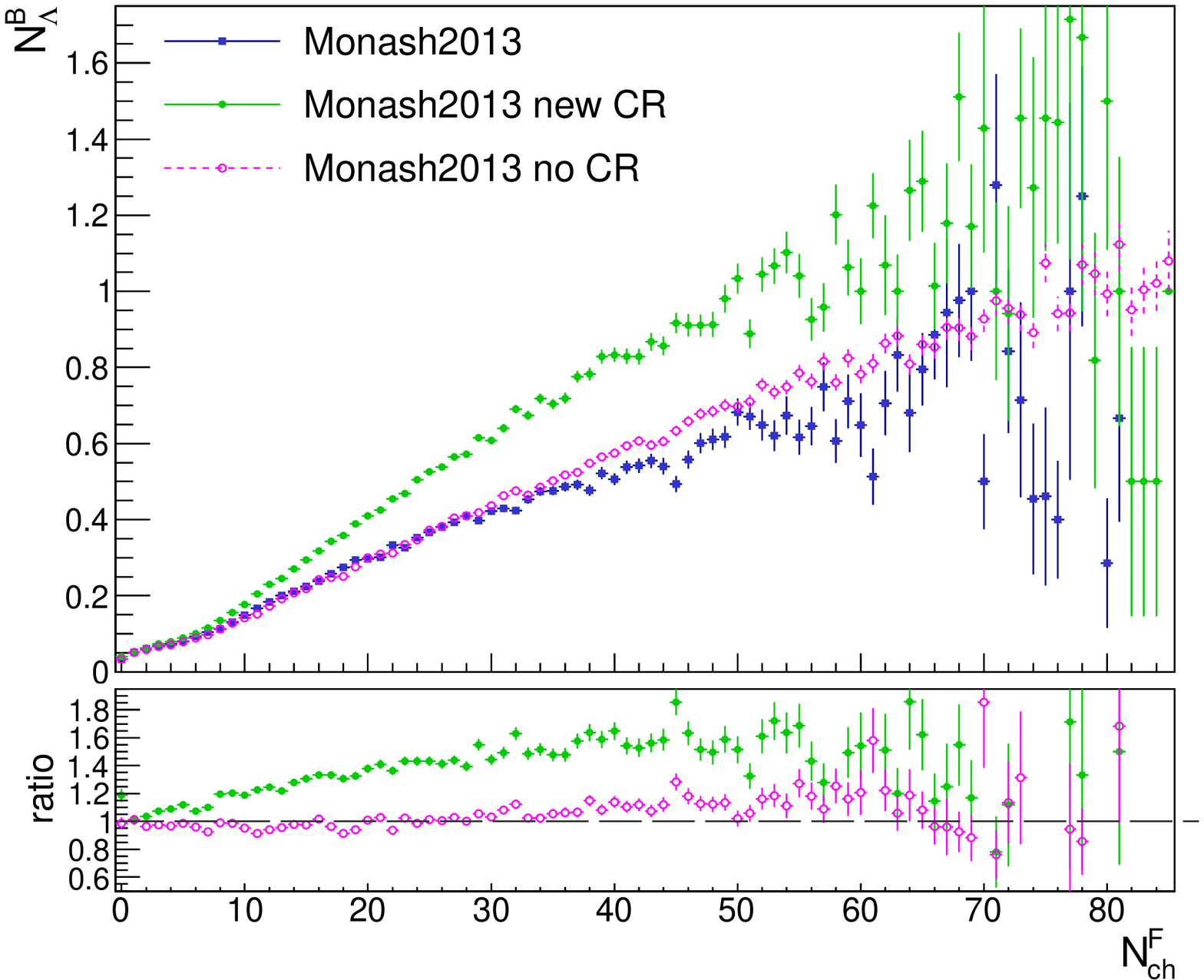} 
 \put(14,60){\small PYTHIA8 }  
\put(14,54){$\av{\Lambda}$ vs $N_{ch}$}
\end{overpic} 
}
\vspace{-0.6cm}
\caption{
(a): $\av{\pt}_{N_{\rm ch}}$ correlations in PYTHIA8 with CR (filled markers) and without CR (open
markers) for a single $eta$-window (blue markers) and the  $-2.4<\eta< -0.5$ and $0.5<\eta< 2.4$
$eta$-windows (red markers).  (b): FB correlation between $\Lambda$/$\overline{\Lambda}$ in backward
window and $N_{\rm ch}$ in forward window, for three configurations of PYTHIA8.
}
\label{fig:PtN_LambdaN_PYTHIA} 
\end{figure}

The effect of MPI and CR on FB correlations were studied in PYTHIA8. 
As shown in Fig.~\ref{fig:PtN_LambdaN_PYTHIA}(a), the $\av{\pt}_{N_{\rm ch}}$ correlation between
single $y$-window is preserved when considering two windows with large rapidity separation.  
When the color reconnection mechanism in PYTHIA8 is switched off the correlation drops to almost
flat behavior as is shown in the Fig.~\ref{fig:PtN_LambdaN_PYTHIA}(a) with open markers.  Using the
FB correlation approach, it is possible to examine  string configurations and their interactions
along $\eta$-range, accessible in an experiment, and also to get rid of short-range contributions
coming from resonance decays, jets etc.

\begin{figure}
\centering
\begin{overpic}[width=0.98\textwidth]
{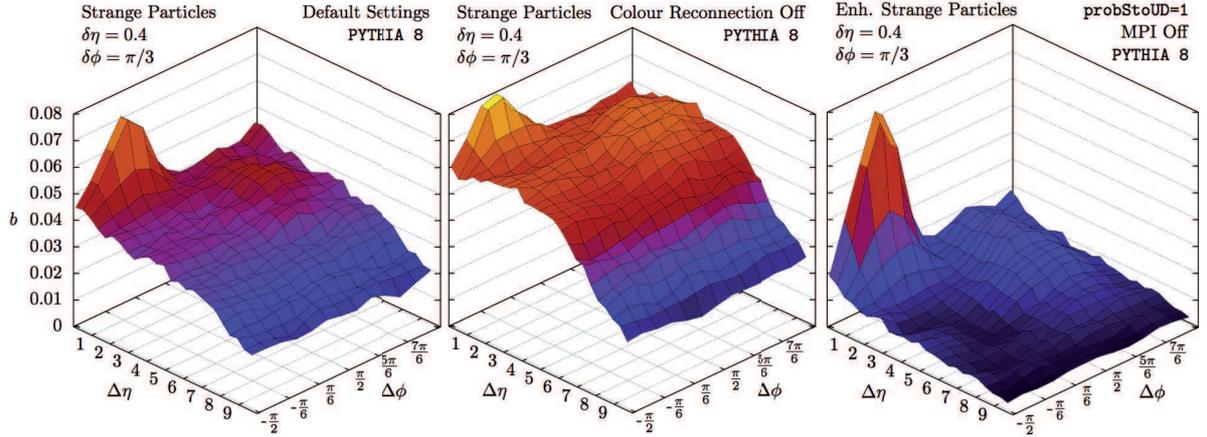} 
\end{overpic}
\vspace{-0.32cm}
\caption{
The strange particle multiplicity correlation topology in $\eta$-$\varphi$ windows in PYTHIA8 with
CR on (a), CR off (b) and with enhanced strangeness production (c).
}
\label{fig:eta-phi_strange} 
\end{figure}

FB correlation involving strange particles can also be used to test string models.
Fig.~\ref{fig:PtN_LambdaN_PYTHIA}(b) shows that the $\av{\Lambda$/$\overline{\Lambda}}_{N_{\rm ch}}$
correlation function is affected by the choice of CR model. 
At $N_{\rm ch}^{\rm F}\approx8$, a ``knee" can be seen in the correlation function. This indicates
some threshold behavior incorporated in PYTHIA8.  
Fig.~\ref{fig:eta-phi_strange} compares the FB strange particle multiplicity correlation in
$\eta$-$\varphi$ windows, obtained in PYTHIA8.  Specifically, it shows that the ``plateau" level and
the shapes of the near-side and away-side structures in this topology changes in the absence of CR.

The effective quark masses used in the Schwinger-mechanism are tuned the $s/u$ ratio seen in experimental data.
In PYTHIA, this implies an explicit suppression of strange quark production, $u : d : s
\approx1 : 1 : 0.3$ \cite{GenPurpEvGen}.  In Fig.~\ref{fig:eta-phi_strange}(c), the 
particle multiplicity correlation topology in $\eta$-$\varphi$ windows is shown for enhanced
strangeness production with MPI turned off, revealing additional modifications of the correlation
coefficient $\bcor$.
Additional figures can be found in attachments for these proceedings.

\section{Conclusions}

A number of observables in $pp$ and AA collisions can not be described by independently hadronizing
particle emitters, indicating the presence of collective effects. 
To study this collectivity, conventional analysis of forward-backward correlations can be extended
from charged particle multiplicities to other observables chosen in the windows in phase-space.
Usage of the ``intensive" variables in FB correlations like $\av{\pt}$,
$\Lambda/\pi$, $K/\pi$, as well as strange  particle yields allows string
interaction mechanisms in $pp$ and AA collisions to be studied.
To properly understand the underlying physics, the shape of the correlation function can be more
informative than using the correlation coefficient $\bcor$ alone.

It was shown that different color reconnection models in PYTHIA8 change the
behavior of FB correlations.
The newest CR scheme in PYTHIA8 gives more baryons and demonstrates different slopes of FB
correlation functions.  The correlations are also affected by MPI and explicit strangeness
suppression in this generator.


\vspace{-0.3cm}
\section*{\it Acknowledgements}
This work is supported for I.A. and G.F. by the Saint-Petersburg State University research grant  11.38.242.2015.

\end{document}